\documentclass[11pt,twoside]{article}
\usepackage{amsmath}
\usepackage{amssymb}
\usepackage{amscd}
\usepackage{mathrsfs}
\usepackage{epstopdf}
\renewcommand{\a}{\alpha}
\renewcommand{\b}{\beta}
\newcommand{\g}{\gamma}
\newcommand{\G}{\Gamma}
\renewcommand{\d}{\delta}
\newcommand{\e}{\varepsilon}
\newcommand{\ze}{{\zeta}}
\renewcommand{\th}{\theta}
\newcommand{\Th}{\Theta}
\newcommand{\io}{\iota}
\renewcommand{\l}{\lambda}
\newcommand{\m}{\mu}
\newcommand{\n}{\nu}
\renewcommand{\r}{\rho}
\newcommand{\s}{\sigma}
\renewcommand{\t}{\tau}
\newcommand{\Cs}{{\rlap{\lower3pt\hbox{\textnormal{\LARGE \char'040}}}{\Gamma}}{}}
\newcommand{\de}{\partial}
\newcommand{\rdg}{{\textstyle\sqrt{{\scriptstyle|}g{\scriptstyle|}}\,}}
\newcommand{\oh}{\tfrac{1}{2}}
\newcommand{\ih}{\tfrac{\iO}{2}}

\newcommand{\osq}{\tfrac{1}{\surd2}}
\newcommand{\isq}{\tfrac{\iO}{\surd2}}
\newcommand{\cj}[1]{\overline{#1}}
\newcommand{\lin}{{\scriptscriptstyle\bigstar}}
\renewcommand{\.}{{\scriptstyle\boldsymbol{\dot{}}}}
\newcommand{\Lll}{{\scriptscriptstyle{\mathrm{L}}}}
\newcommand{\Rrr}{{\scriptscriptstyle{\mathrm{R}}}}
\newcommand{\interaction}{{}_{\sst{\mathrm{int}}}}
\newcommand{\sbot}{{\scriptscriptstyle\bot}}
\newcommand{\sbo}{{\!\sbot}}
\newcommand{\bl}{{\bar\lambda}}
\newcommand{\bm}{{\bar\mu}}
\newcommand{\bs}{{\bar s}}
\newcommand{\bu}{{\bar u}}
\newcommand{\bv}{{\bar v}}
\newcommand{\bw}{{\bar w}}
\newcommand{\be}{{\bar\varepsilon}}
\newcommand{\bze}{{\bar\zeta}}
\newcommand{\bch}{{\bar\chi}}
\newcommand{\E}{{\boldsymbol{E}}}
\newcommand{\F}{{\boldsymbol{F}}}
\newcommand{\Fl}{\F^\lin}
\newcommand{\Fc}{\cj{\F}}
\newcommand{\Fa}{\cj{\F}{}^\lin}
\newcommand{\FLa}{\Fa_{\!\!\Lll}}
\newcommand{\FRa}{\Fa_{\!\!\Rrr}}
\newcommand{\Gb}{{\boldsymbol{G}}}
\newcommand{\GA}{{\boldsymbol{\Gamma}}}
\newcommand{\GAG}{{\boldsymbol{\Gamma}}_{{\!}_\Gb}}
\renewcommand{\H}{{\boldsymbol{H}}}
\newcommand{\I}{{\boldsymbol{I}}}
\newcommand{\Ic}{\cj{\I}}
\newcommand{\Il}{\I{}^\lin}
\newcommand{\M}{{\boldsymbol{M}}}
\newcommand{\Pm}{{\boldsymbol{P}}_{\!\!m}}
\let\Sec=\S
\renewcommand{\S}{{\boldsymbol{S}}}
\newcommand{\Sc}{\cj{\S}}
\newcommand{\Sa}{\cj{\S}{}^\lin}
\newcommand{\Sl}{\S{}^\lin}
\newcommand{\U}{{\boldsymbol{U}}}
\newcommand{\Uc}{\cj{\U}}
\newcommand{\Ua}{\cj{\U}{}^\lin}
\newcommand{\Ul}{\U{}^\lin}
\newcommand{\V}{{\boldsymbol{V}}}
\newcommand{\Vc}{\cj{\V}}
\newcommand{\W}{{\boldsymbol{W}}}
\newcommand{\Wc}{\cj{\W}}
\newcommand{\Wa}{\cj{\W}{}^\lin}
\newcommand{\Wl}{\W{}^\lin}
\newcommand{\Y}{{\boldsymbol{Y}}}
\newcommand{\Yc}{\cj{\Y}}
\newcommand{\Ya}{\cj{\Y}{}^\lin}
\newcommand{\Yl}{\Y{}^\lin}
\newcommand{\Lie}{\mathfrak{L}}
\newcommand{\Ug}{\boldsymbol{\mathrm{U}}}
\newcommand{\CC}{{\mathbb{C}}}
\newcommand{\LL}{{\mathbb{L}}}
\newcommand{\RR}{{\mathbb{R}}}
\newcommand{\ZZ}{{\mathbb{Z}}}
\newcommand{\Hcal}{{\mathcal{H}}}
\newcommand{\Ical}{{\mathcal{I}}}
\newcommand{\Jcal}{{\mathcal{J}}}
\newcommand{\Lcal}{{\mathcal{L}}}
\newcommand{\Ocal}{{\mathcal{O}}}
\newcommand{\Scal}{{\mathcal{S}}}
\newcommand{\lfr}{\mathfrak{l}}
\newcommand{\OC}{{\boldsymbol{\Ocal}}}
\newcommand{\End}{\operatorname{End}}
\newcommand{\Aut}{\operatorname{Aut}}
\newcommand{\Tr}{\operatorname{Tr}}
\newcommand{\Ker}{\operatorname{Ker}}
\newcommand{\Id}[1]{{1\!\!1}\!{}_{#1}{}}
\newcommand{\id}{{1\!\!1}}
\newcommand{\dO}{\mathrm{d}}
\newcommand{\DO}{\mathrm{D}}
\newcommand{\HO}{\mathrm{H}}
\newcommand{\JO}{\mathrm{J}}
\newcommand{\TO}{\mathrm{T}}
\newcommand{\TS}{\TO^{*}\!}
\newcommand{\iO}{\mathrm{i}}
\newcommand{\na}{\nabla\!}
\newcommand{\ten}[1]{\operatorname*{\otimes}_{\!{\scriptscriptstyle #1}} }
\newcommand{\cart}[1]{\operatorname*{\times}_{\!{\scriptscriptstyle #1}} }
\newcommand{\dir}[1]{\operatorname*{\oplus}_{\!{\scriptscriptstyle #1}} }
\newcommand{\we}{{\,\wedge\,}}
\newcommand{\weu}[1]{{\wedge^{\!#1}}}

\newcommand{\comp}{\mathbin{\raisebox{1pt}{$\scriptstyle\circ$}}}
\newcommand{\tn}{{\,\otimes\,}}
\newcommand{\bwe}{\,{\barwedge}\,}
\newcommand{\sco}[3]{\mathsf{c}\Ii{#1}{#2#3}}
\newcommand{\bang}[1]{{\langle#1\rangle}}
\newcommand{\Ii}[2]{{}^{#1}_{\phantom{#1}\!#2}}
\newcommand{\iI}[2]{{}_{#1}^{\phantom{#1}\!#2}}
\newcommand{\iIi}[3]{{}_{#1\phantom{#2}\!\!#3}^{\phantom{#1}\!#2}}
\newcommand{\sA}{{\scriptscriptstyle A}}
\newcommand{\sB}{{\scriptscriptstyle B}}
\newcommand{\sC}{{\scriptscriptstyle C}}
\newcommand{\sD}{{\scriptscriptstyle D}}
\newcommand{\cA}{{\sA\.}}
\newcommand{\cB}{{\sB\.}}
\newcommand{\cC}{{\sC\.}}
\newcommand{\cD}{{\sD\.}}
\newcommand{\AAd}{{\sA\cA}}
\newcommand{\BBd}{{\sB\cB}}
\newcommand{\CCd}{{\sC\cC}}
\newcommand{\DDd}{{\sD\cD}}
\newcommand{\ee}{{\mathsf{e}}}
\newcommand{\sref}[1]{\Sec\ref{#1}}
\newcommand{\ie}{i.e$.$}
\newcommand{\eg}{e.g$.$}
\newcommand{\sst}{\scriptscriptstyle}
\newcommand{\onto}{\rightarrowtail}
\newcommand{\YL}{\Y\!_{\!\Lll}} \newcommand{\YR}{\Y\!_{\!\Rrr}}
\newcommand{\YLc}{\Yc\!_{\!\Lll}} \newcommand{\YRc}{\Yc\!_{\!\Rrr}}
\newcommand{\FL}{\F_{\!\!\Lll}} \newcommand{\FR}{\F_{\!\!\Rrr}}
\newcommand{\FLc}{\Fc_{\!\!\Lll}} \newcommand{\FRc}{\Fc_{\!\!\Rrr}}
\newcommand{\FLl}{\Fl_{\!\!\Lll}} \newcommand{\FRl}{\Fl_{\!\!\Rrr}}
\newcommand{\reH}{{\scriptstyle\Hcal}}
\newcommand{\Hivac}{{\scriptstyle\Hcal}_{\sst0}}
\newcommand{\bHivac}{\bar{\scriptstyle\Hcal}_{\sst0}}
\newcommand{\phiplus}{\phi_{{}_{+}}}
\newcommand{\phiminus}{\phi_{{}_{-}}}
\newcommand{\phizero}{\phi_{{}_{0}}}
\newcommand{\gh}{\omega}
\newcommand{\ghR}{\gh_\Rrr} \newcommand{\ghL}{\gh_\Lll}
\newcommand{\agh}{\varpi}
\newcommand{\aghR}{\agh_\Rrr} \newcommand{\aghL}{\agh_\Lll}
\newcommand{\FH}{\Omega}
\title{Natural extensions of electroweak geometry \\
and Higgs interactions}
\date{{\small December 4, 2014} }
\author{Daniel Canarutto\\[6pt]
{\small\it Dipartimento di Matematica e Informatica ``U.~Dini'', }\\
{\small\it Via S. Marta 3, 50139 Firenze, Italia}\\
{\small email:~daniel.canarutto@unifi.it}\\
{\small http://www.dma.unifi.it/\char126 canarutto}}
\begin{document}
\maketitle
\begin{abstract}\noindent
We explore the possibility that the Higgs boson of the standard model
be actually a member of a larger family,
by showing that a more elaborate internal structure
naturally arises from geometrical arguments,
in the context of a partly original handling of gauge fields
which was put forward in previous papers.
A possible mechanism yielding the usual Higgs potential is proposed.
New types of point interactions,
arising in particular from two-spinor index contractions,
are shown to be allowed.
\end{abstract}

\bigbreak

\bigbreak\noindent

\bigbreak
\noindent
{\sc Journal reference}:\\
D.\ Canarutto:
`Natural extensions of electroweak geometry and Higgs interactions',\\
Annales Henri Poincar\'e (2014),
DOI: 10.1007/s00023-014-0383-8 \\
Springer link: http://link.springer.com/article/10.1007/s00023-014-0383-8

\tableofcontents

\section*{Introduction}

The standard model does not fix all the properties of the Higgs boson,
nor does experiment.
Accordingly, various extensions and variations of the existing theory
have been explored~\cite{LM,De02,Ta,Fa,FKD,RySh,MT,Mo10,NoBi,AMS}.
Recent results give us hope that fresh experimental evidence
may allow finer discrimination among physical theories.

In this paper we'll undertake a further exploration
and show that a possible larger internal structure of electroweak geometry
naturally arises from geometrical arguments,
in the context of a partly original handling of gauge fields
which was described in previous papers~\cite{C10a,C12a}.
We start from the observation that a matter field,
in classical field theory, is described as a section of some bundle over spacetime,
while a gauge field is a connection of that bundle.
Now a linear connection, possibly preserving some fibered algebraic structure,
can also be seen as a section of a finite-dimensional bundle,
but not of a vector bundle.
A field to be quantised, however,
\emph{must} be a section of a vector bundle,
as one obtains the related \emph{quantum bundle} via tensor product
by a certain (infinite-dimensional) $\ZZ_2$-graded algebra~\cite{C14b};
the need for gauge-fixing can be viewed as stemming from this difficulty.

This author's views about such issues
tend towards a somewhat radical ansatz\footnote{%
Described in a short essay~\cite{C11a}
partly inspired by ideas of Penrose~\cite{Penrose71}.} %
of the kind `the system defines the geometry'.
As a provisional, middle-of-the-way attitude,
here we explore an approach in which all integer-spin bundles
arise as tensor products of half-spin bundles,
and interactions are described by fiber contractions,
possibly in all allowed ways.
The underlying idea is that the various tensor factors could be seen
as roughly analogue to `chemical bonds'.
Accordingly, we view gauge fields and connections as different though related notions,
and regard connections as background `macroscopic' structures,
on the same footing of fields describing a fixed gravitational background.
The relation between gauge fields and connections depends on gauge freedom,
which can be seen to arise in a natural way
from the geometry of two-spinors (Weyl spinors).
This point of view allows new interactions to be considered,
which in general do not preserve gauge symmetry in a strict sense;
gauge symmetry is still preserved, however, by the standard interactions.

The above said procedure for producing integer-spin sectors
actually turns out to yield all the sectors of the standard electroweak theory,
and some more.
One may object that the Standard Model
looks already sufficiently complicated as it is;
only experiment, however, can eventually tell us what to drop.
In particular, we'll see that one new sector is involved in a proposed mechanism
for recovering the Higgs potential, which is usually inserted `by hand';
we suggest that this mechanism is related to the question
of `breaking of dilatonic symmetry'.

We also exhibit various further interactions, related to 2-spinor geometry,
in the context of an extended Higgs sector which arises again
from our general procedure.
Indeed, 2-spinors can be regarded as fundamental building blocks,
rather than just the basis of a useful formalism.

\section{A geometric setting for gauge field theories}
\label{s:A geometric setting for gauge field theories}
\subsection{Remarks about classical and quantum gauge theories}
\label{ss:Remarks about classical and quantum gauge theories}

We begin by expanding 
some of the preliminary obervations made in the introduction.
A `matter field' in a classical field theory
is a section of a vector bundle \hbox{$\E\onto\M$}
over the spacetime manifold $\M$.
Linear connections can be seen as sections of an affine bundle \hbox{$\GA\onto\M$}.
Actually, denoting by \hbox{$\JO\E\onto\E$} the \emph{first jet bundle},
we find that \hbox{$\GA\subset\JO\E\ten{\M}\E^*\onto\M$}
is the affine sub-bundle projecting over the identity $\Id{\E}$\,.
Its derived vector bundle
(the bundle of `differences of linear connections')
is \hbox{$\DO\GA=\TS\M\ten{\M}\End\E$}\,.

Symmetries in such theories are usually treated in terms of matrix groups
and principal bundles.
Let me propose a somewhat different (though eventually equivalent) description.
First, note that \hbox{$\End\E\cong\E\ten{\M}\E^*$}
has the subbundle $\Aut\E$ over $\M$ whose fibers are the groups
of all fiber automorphisms (\ie\ invertible endomorphisms),
and that the fibers of $\End\E$,
with the product given by the ordinary commutator,
are the Lie algebras of the fibers of $\Aut\E$.
Now suppose that the fibers of $\E$ are smootly endowed
with some algebraic structure;
this selects the sub-bundle \hbox{$\Gb\subset\Aut\E$}
whose fibers are constituted by all automorphisms
\emph{which preserve that structure}.
This is a possibly non trivial \emph{Lie group bundle},
whose Lie algebra bundle is a sub-bundle \hbox{$\Lie\subset\End\E$}.
If we restrict ourselves to consider connections which make
the fiber structure covariantly constant,
then the difference of any two such connections is $\Lie$-valued.
Accordingly, such a connection is a section of an affine sub-bundle
\hbox{$\GAG\subset\GA$}\,,
whose derived vector bundle is \hbox{$\DO\GAG=\TS\M\ten{\M}\Lie$}\,.
Locally, we recover the usual matrix formalism by fixing
any special frame of $\E$,
that is a frame which is `adapted' to the fiber structure.

The basic example is that when $\E$
is a complex vector bundle with a Hermitian scalar product in the fibers;
then the fibers of $\Lie$ are constituted by all anti-Hermitian endomorphisms,
and the special frames apt to simplify calculations are the orthonormal frames.
Another important case is that of spinor bundles~\cite{C07};
in the Dirac bundle, in particular, one uses Weyl and Dirac frames.
Similar considerations can be made about the related bundle \hbox{$\H\onto\M$},
introduced
in~\sref{ss:Two-spinors and Einstein-Cartan-Maxwell-Dirac field theory},
whose fibers are naturally endowed with a Minkowskian structure,
as well as about the tangent bundle $\TO\M$.

As a previous paper~\cite{C14b} discussed to some extent,
one can only quantise classical fields which are sections of a vector bundle,
since the procedure requires constructing a new `quantum bundle'
obtained from the classical bundle by tensorializing its fibers
by a certain `operator space' $\OC$
(a $\ZZ_2$-graded infinite-dimensional algebra).
Hence one resorts to \emph{gauge fixing},
namely the local choice of a flat connection $\g_0$\,;
any other connection $\g$ is then characterized by the difference
\hbox{$\a\equiv\g-\g_0:\M\to\TS\M\ten{\M}\Lie$}\,.

According to the standard procedure, one derives particle interactions
from the various terms in the Lagrangian density
of the theory under consideration.
In momentum formulation one directly recovers the interactions
by replacing the curvature tensor of $\g$ with the ``curvature-like'' tensor
$$\r[\a]:=\iO\,p\we\a+\a\bwe\a~,$$
where $p$ is the gauge boson's momentum and the notation $\a\bwe\b$
stands for exterior product of $\Lie$-valued forms together with composition
(if $\a$ and $\b$ are $\Lie$-valued 1-forms
then \hbox{$(\a\bwe\b)\iI{ab}i=\sco ijk\a\iI aj\,\b\iI bk$} where
\hbox{$\sco ijk\equiv[\lfr_j\,,\,\lfr_k]^i$}
are the `structure constants' in the chosen special frame
$\bigl(\lfr_i\bigr)$ of $\Lie$).
We observe that the replacement \hbox{$\a\to p\tn\chi+\a$}\,,
with \hbox{$\chi:\M\to\Lie$}\,, preserves $\r(\a)$\,.
Moreover by examining point interactions in terms of 2-spinors~\cite{C14a}
one can show that such replacement
does not affect scattering matrix calculations.
Hence the physical meaning of the gauge field is encoded
in its equivalence class,
$\a$ and $\a'$ being equivalent if their difference is of the kind $p\tn\chi$\,.

On the other hand,
a particle's momentum is strictly related to its spin,
so in a sense (to be made more precise) we may see it as an `internal' property,
thus inviting us to view all possible particle interactions
as dictated by contractions among tensor factors in the fibers.
Indeed, the Lagrangian density itself is essentially dictated
by the underlying bundle geometry;
viewing the `chemical bond' approach as fundamental,
we are incouraged to take all contractions into account.
Eventually, of course, we'll be able to include all of them
in a suitable extension of the usual Lagrangian.
Furthermore, the possible arrangements of involved tensor factors
may yield natural extensions of the usually considered sectors.

The role of a classical connection, in the above sketched scheme,
may be thought of as that of a background macroscopic structure,
analogous to gravity.

\subsection[Two-spinors and ECMD field theory]%
{Two-spinors and Einstein-Cartan-Maxwell-Dirac field theory}
\label{ss:Two-spinors and Einstein-Cartan-Maxwell-Dirac field theory}

A partly original approach to 2-spinors,
discussed in previous papers~\cite{C98,C00b,C07},
turns out to be convenient
for an integrated treatment of classical Einstein-Cartan-Maxwell-Dirac fields
starting from minimal geometric assumptions.
We first summarize the basic algebraic results.

If $\V$ is a complex vector space and $\Vc$ is its conjugate space,
then Hermitian transposition is a natural anti-linear involution of $\V\tn\Vc$,
which can be decomposed into the direct sum
of its Hermitian and anti-Hermitian \emph{real} subspaces.
Starting from a $2$-dimensional complex vector space $\S$,
with no further assumption,
one gets a rich algebraic structure:
\smallbreak\noindent
$\bullet~$%
The Hermitian subspace of \hbox{$\weu2\S\tn\weu2\Sc$}
is a real $1$-dimensional vector space with a distinguished orientation;
its positively oriented semispace $\LL^2$
(whose elements are of the type $w\tn\bw$\,, $w\in\weu2\S$)
has the square root semispace $\LL$,
which can be identified with the space
of \emph{length units}.\footnote{%
For a review of unit spaces see \eg\ \cite{C00b,JMV10,C12a}.} %
\smallbreak\noindent
$\bullet~$%
The $2$-spinor space is defined to be $\U:=\LL^{-1/2}\tn\S$.
The space $\weu2\U$ is naturally endowed with a Hermitian metric,
namely the identity element in\footnote{ 
We distinguish between the \emph{complex dual},
indicated by the symbol $\lin$, and the \emph{real dual},
indicated by a regular asterisk $*$.} 
$$\HO[(\weu{2}\Ua)\tn(\weu{2}\Ul)]
\cong\LL^2\tn\HO[(\weu{2}\Sa)\tn(\weu{2}\Sl)]~,$$
so that normalized `symplectic forms' $\e\in\weu2\Ul$
constitute a $\Ug(1)$-space\footnote{%
Differently from the usual $2$-spinor formalism,
no symplectic form is held fixed.
Also note that no Hermitian form on $\S$ or $\U$ is assigned;
actually, because of the Lorentz structure of $\H$ (below),
the choice of such an object turns out to be equivalent
to the choice of an `observer'.} %
(any two of them are related by a phase factor).
Each $\e$ yields the isomorphism
\hbox{$\e^\flat:\U\to\Ul:u\mapsto u^\flat:=\e(u,\_)$}\,.
\smallbreak\noindent
$\bullet~$%
The identity element in $\HO[(\weu{2}\Ua)\tn(\weu{2}\Ul)]$
can be written as $\e\tn\be$ where $\e\in\weu2\Ul$
is any normalized element.
This natural object can also be seen as a bilinear form $g$ on $\U\tn\Uc$,
via the rule
\hbox{$g(u\tn\bv,r\tn\bs)=\e(u,r)\,\be(\bv,\bs)$}
extended by linearity.
Its restriction to the Hermitian subspace $\H\equiv\HO(\U\tn\Uc)$
turns out to be a Lorentz metric.
Null elements in $\H$ are of the form $\pm u\tn\bu$ with $u\in\U$
(thus there is a distinguished time-orientation in $\H$).
\smallbreak\noindent
$\bullet~$Let $\W\equiv\U\oplus\Ua$.
The linear map
$\g:\U\tn\Uc\to\End(\W):y\mapsto\g(y)$ characterized by
$$\tilde\g(r\tn\bs)(u,\chi)
=\sqrt2\bigl(\bang{\chi,\bs}\,r\,,\bang{r^\flat,u}\,\bs^\flat\,\bigr)$$
is well-defined independently of the choice of the normalized $\e\in\weu2\Ul$
yielding the isomorphism $\e^\flat$.
Its restriction to $\H$ turns out to be a Clifford map.
Thus one is led to regard $\W\equiv\U\oplus\Ua$ as the space of Dirac spinors,
decomposed into its Weyl subspaces.
The anti-isomorphism $\W\to\Wl:(u,\chi)\mapsto(\bch,\bu)$
is the usual \emph{Dirac adjunction}
(\hbox{$\psi\mapsto\bar\psi$} in traditional notation),
associated with a Hermitian product having the signature $({+},{+},{-},{-})$\,.

We now consider a complex vector bundle $\S\onto\M$
with $2$-dimensional fibers.
By performing the above sketched constructions fiberwise
we obtain various vector bundles, which are denoted, for simplicity,
by the same symbols.
We observe that some appropriate topological restrictions
are implicit in what follows;
we'll assume the needed hypotheses to hold without further comment.

A linear connection $\Cs$ on $\S$ determines linear connections
on the associated bundles, and, in particular, connections
$G$ of $\LL$, $X$ of $\weu2\U$ and $\tilde\G$ of $\H$;
on turn, it can be expressed in terms of these as
$$\Cs\iIi{a}{\sA}{\sB}=(G_a+\iO\,X_a)\d\Ii{\sA}{\sB}
+\oh\,\tilde\G\iIi{a}{\sA\cA}{\sB\cA}$$
(dotted indices refer to components in conjugate spaces).

If $\M$ is $4$-dimensional, then a \emph{tetrad}
(or a \emph{soldering form})
is defined to be a linear morphism $\Th:\TO\M\to\LL\tn\H$.
An invertible tetrad determines, by pull-back,
a Lorentz metric on $\M$ and a metric connection of $\TO\M\onto\M$,
as well as a Dirac morphism $\TO\M\to\LL\tn\End\W$.

A non-singular field theory in the above geometric environment
can be naturally formulated~\cite{C98}
even if $\Th$ is not required to be invertible everywhere.
If the invertibility requirement is satisfied
then one gets essentially the standard Einstein-Cartan-Maxwell-Dirac theory,
but with some redefinition of the fundamental fields:
these are now the $2$-spinor connection $\Cs$,
the tetrad $\Th$,
the Maxwell field $F$ and the Dirac field $\psi:\M\to\LL^{-3/2}\tn\W$.
Gravitation is represented by $\Th$ and $\tilde\G$ together.
$G$ is assumed to have vanishing curvature,
$\dO G=0$, so that we can find local charts such that $G_a=0$\,;
this amounts to `gauging away' the conformal `dilaton' symmetry.
Coupling constants arise as covariantly constants sections of $\LL^r$
($r$ rational).
One then writes a natural Lagrangian which yields all the field equations:
the Einstein equation and the equation for torsion;
the equation \hbox{$F=2\,\dO X$}, whence \hbox{$\dO F=0$}
(thus $X$ is essentially the electromagnetic potential),
and the other Maxwell equation;
finally, one gets the Dirac equation~\cite{C00b}.

\subsection{On gauge freedom in QED}
\label{ss:On gauge freedom in QED}

We work in a context where the gravitational field is treated
as a fixed background structure;
this means that the tetrad $\Th$ and the gravitational part of the
spin connection are fixed (rather than `field variables').
If no confusion arises,
by using $\Th$ we make the identification \hbox{$\TO\M\cong\LL\tn\H$},
and view 1-forms of $\M$ as \emph{scaled} sections
$\M\to\LL^{-1}\tn\H^*$.

Besides the Weyl decomposition,
in QED one needs the Dirac decomposition.
Consider the subbundle $\Pm\subset\TS\M$ over $\M$
whose fibers are the future hyperboloids (`mass-shells')
corresponding to mass $m\in\{0\}\cup\LL^{-1}$.
If $p\in(\Pm)_x$\,, $x\in\M$, then we have
$$\W\!\!_x=\W^{+}_{\!\!p}\oplus\W^{-}_{\!\!p}~,\quad
\W^{\pm}_{\!\!p}:=\Ker(\g[p^\#]\mp m)~,$$
where $p^\#\equiv g^\#(p)\in\LL^{-2}\tn\TO\M$
is the contravariant form of $p$\,.
Thus we obtain 2-fibered bundles $\W_{\!\!m}^\pm\onto\Pm\onto\M$,
called the \emph{electron bundle} and the \emph{positron bundle}, where
$$\W_{\!\!m}^\pm:=\bigsqcup_{p\in\Pm}\W_{\!\!p}^\pm\subset\Pm\cart{\M}\W~.$$
This splitting has an interesting two-spinor description~\cite{C07}.
If \hbox{$\psi\equiv(u,\bl)\in\W$} then
$$\t\equiv\tfrac1{\sqrt2\,|\bang{\l,u}|}\,(u\tn\bu+\l^\#\tn\bl^\#)\in\H$$
is a unit future-pointing timelike vector
(here \hbox{$\l^\#\equiv\e^\#(\l)$}
where $\e^\#$ is the inverse isomorphism of $\e^\flat$,
see~\sref{ss:Two-spinors and Einstein-Cartan-Maxwell-Dirac field theory}).
By a straightforward calculation one sees that \hbox{$\g[\t]\psi=\pm\psi$}
if and only if \hbox{$\bang{\l,u}\in\RR^\pm$}.
Conversely, it can be proved that if \hbox{$\t'\in\H$} is such that
\hbox{$\g[\t']\psi=\pm\psi$}\,,
then necessarily \hbox{$\t'=\t$}\,.
It follows that internal states of free electrons and positrons
carry the full information about their momenta.

The interaction between the Dirac field and the e.m.\ potential
can be deduced from the Dirac Lagrangian
by writing $X=e\,A$\,, where $A$ is a true 1-form, via the choice of an e.m.\ gauge.
Somewhat differently we can see the interaction as directly deriving
from the underlying geometric structure,
namely as the natural contraction
$$\ell\interaction:\Wc\tn\H\tn\W\to\CC:
(\bar\psi,A,\psi')\mapsto -e\,\bang{\bar\psi,\g[A]\psi'}~.$$
We can also see $\ell\interaction$ as a tensor field $\M\to\Wa\tn\H^*\tn\Wl$.
From this,
using the algebraic structures of the fibers of the involved bundles,
we can obtain eight tensor fields of different index types;
these correspond to different combinations
of particle absorption and emission,
respectively represented by covariant and contravariant indices.

In two-spinor terms, if
\hbox{$\psi=(v,\bm)$} and \hbox{$\psi'=(u,\bl)$} we get
\begin{align*}
\bang{\bar\psi,\g[r\tn\bs]\psi'}&=
\sqrt2\,\bigl(\bang{\m,r}\,\bang{\bl,\bs}+\e(u,r)\,\be(\bv,\bs)\bigr)\equiv
\\[6pt]
&\equiv\sqrt2\,g\bigl(r\tn\bs\;,\;u\tn\bv+\m^\#\tn\bl^\#\bigr)~,
\end{align*}
hence in general
\hbox{$\bang{\bar\psi,\g[A]\psi'}=
\sqrt2\,g\bigl(A^\#,u\tn\bv+\m^\#\tn\bl^\#\bigr)$}\,.
The kernel of the map
$$\bang{\bar\psi,\g[\_]\psi'}:
\CC\tn\H^*\to\CC:A\mapsto\bang{\bar\psi,\g[A]\psi'}$$
is then constituted by all covectors orthogonal to
$$u\tn\bv+\m^\#\tn\bl^\#\in\U\tn\Uc=\CC\tn\H~.$$

By 2-spinor algebra calculations~\cite{C14a} one then checks
that the interaction is unaffected by adding the algebraic sum
of the interacting fermions' momenta to the internal photon state.
As observed in~\sref{ss:Remarks about classical and quantum gauge theories},
from a certain point of view this fact can be seen as the basis of gauge freedom.
Furthermore, the above argument is easily extended to non-abelian theories
in the setting we are going to describe.

\subsection{Two-spinors and non-abelian gauge theories}
\label{ss:Two-spinors and non-abelian gauge theories}

We generalize the fermion bundle of electrodynamics
(\sref{ss:Two-spinors and Einstein-Cartan-Maxwell-Dirac field theory},%
\;\ref{ss:On gauge freedom in QED})
by allowing an extended fiber structure
which may violate the symmetry between right-handed and left-handed sectors.
The new fermion bundle \hbox{$\Y\onto\M$} is given by
$$\Y\equiv\YR\oplus\YL\equiv(\FR\tn\U)\oplus(\FL\tn\Ua)~,$$
where $\F_{\!\!\Rrr}\onto\M$ and $\F_{\!\!\Lll}\onto\M$
are complex vector bundles
(describing the internal fermion structure besides spin)
endowed with fibered Hermitian structures.

According to our view
(\sref{ss:Remarks about classical and quantum gauge theories}),
particle interactions should be related to contractions in the fibers;
we assume that the fundamental bosons,
rather than being directly derived from connections
of the fermion bundle sectors,
are described as sections of vector bundles arising from expanding $\Yc\tn\Y$,
namely
\begin{align*}
\Yc\tn\Y&\cong
(\YRc\tn\YR)\oplus(\YLc\tn\YL)~\oplus~(\YRc\tn\YL)~\oplus~(\YLc\tn\YR)~\cong~{}
\\[6pt]
&\cong~
(\FRc\tn\FR\tn\Uc\tn\U)~\oplus~
(\FLc\tn\FL\tn\Ul\tn\Ua)~\oplus~{}\\[4pt]
&\qquad\qquad{}\oplus~(\FRc\tn\FL\tn\Uc\tn\Ua)~\oplus~
(\FLc\tn\FR\tn\Ul\tn\U)~.
\end{align*}
The standard electroweak interactions will be seen
(\sref{ss:Standard electroweak geometry rivisited})
to arise naturally from this scheme.

Next we observe that the Hermitian structures of $\FR$ and $\FL$
determine fibered isomorphisms
\hbox{$\FRc\cong\FRl$} and \hbox{$\FLc\cong\FLl$}\,,
and by definition we have
$$\U\tn\Uc\cong\CC\tn\H~,\quad \Ul\tn\Ua\cong\CC\tn\H^*~.$$
Furthermore,
the Lorentz metric yields the isomorphism $\H\leftrightarrow\H^*$,
and the tetrad $\Th$ yields the scaled isomorhism
\hbox{$\H^*\leftrightarrow\LL\tn\TS\M$}.
Hence, after rearranging the order of some tensor factors,
sections $\M\to\LL^{-1}\tn\YRc\tn\YR$ and $M\to\LL^{-1}\tn\YLc\tn\YL$
can be seen as fields
\begin{align*}
&\M\to\TS\M\tn\FR\tn\FRl\cong\TS\M\tn\End\FR~,
\\[6pt]
&\M\to\TS\M\tn\FL\tn\FLl\cong\TS\M\tn\End\FL~,
\end{align*}
respectively,
and are obvious candidates for the role of gauge fields
provided we restrict the targets
to $\TS\M\tn\Lie_\Rrr$ and $\TS\M\tn\Lie_\Lll$\,,
where \hbox{$\Lie_\Rrr\subset\End\FR$} and
\hbox{$\Lie_\Lll\subset\End\FL$} are the subbundles
constituted by anti-Hermitian endomorphisms.

The case of sections
$$\M\to\YRc\tn\YL\cong\FRc\tn\FL\tn\Uc\tn\Ua~,\quad
\M\to\YLc\tn\YR\cong\FLc\tn\FR\tn\Ul\tn\U$$
is somewhat different.
We have
$$\Uc\tn\Ua\cong\End\Uc~,\quad\Ul\tn\U\cong\End\U~,$$
so that we can consider, in particular,
those sections which are proportional to the identity of $\Uc$ or $\U$\,;
the Higgs field of the electroweak theory can be seen to arise exactly
in this way (\sref{ss:Standard electroweak geometry rivisited}).
Further possibilities arise from considering
all fibered tensor products between any two of the bundles
$\Y$, $\Yc$, $\Yl$ and $\Ya$, though of course various isomorphisms
limit the number of independent cases.
If we follow this thread,
eventually we may have to deal with several more particle types
and interactions
than would be implied by a na\"{i}f application of ``quantization rules''
based on the underlying classical-field structure,
in which the fundamental bosons are derived from a connection.
In particular, from $\Y\tn\Yl$ and $\Yc\tn\Ya$ we get sectors
$$\End\FR\tn\End\U\,,~\End\FL\tn\End\U\,,~
\End\FR\tn\End\Uc\,,~\End\FL\tn\End\Uc\,,$$
whose subbundles proportional to $\Id{\U}$ and $\Id{\Uc}$
can play the role of ghost and anti-ghost bundles.
\smallbreak
Summarizing, we can identify sections of the above said sectors as follows:
\smallbreak\noindent
$\bullet$~the \emph{matter field} is a section
\hbox{$\Psi\equiv(\Psi_\Rrr\,,\,\Psi_\Lll):\M\to\Y\equiv\YR\oplus\YL$}\,;
\smallbreak\noindent
$\bullet$~\emph{gauge fields} are sections
\hbox{$A_{\Rrr}:\M\to\TS\M\tn\Lie_\Rrr$} and
\hbox{$A_{\Lll}:\M\to\TS\M\tn\Lie_\Lll$}\,;
\smallbreak\noindent
$\bullet$~the \emph{extended Higgs and anti-Higgs fields} are sections
$$\Phi:\M\to\FL\tn\FRl\tn\End\Uc~,\quad
\bar\Phi:\M\to\FR\tn\FLl\tn\End\U~;$$
in the sub-sectors proportional to the identity we write
\hbox{$\Phi\equiv\phi\tn\Id{\Uc}$}
and \hbox{$\bar\Phi\equiv\bar\phi\tn\Id{\U}$}\,,
with $\phi$ and $\bar\phi$ playing the role
of the usual Higgs and anti-Higgs fields

\smallbreak\noindent
$\bullet$~\emph{ghosts} are sections \hbox{$\ghR:\M\to\Lie_\Rrr$} and
\hbox{$\ghL:\M\to\Lie_\Lll$}\,;
\smallbreak\noindent
$\bullet$~\emph{anti-ghosts} are sections \hbox{$\aghR:\M\to\Lie_\Rrr^*$} and
\hbox{$\aghL:\M\to\Lie_\Lll^*$}\,.
\smallbreak\noindent
Ghosts and anti-ghosts precisely
describe the notion of ``infinitesimal gauge transformations''.
They are mutually \emph{independent} fields,
as one sees at once from the different ways in which they appear in the
ghost Lagrangian~\cite{C14b},
the natural isomorphisms \hbox{$\Lie\cong\Lie^*$} notwithstanding.

We could also consider extensions of ghost and anti-ghosts
similar to the extensions introduced in the case of Higgs fields.
Then we would see the above as the particular cases $\ghL\tn\id$ and the like.

Finally, we note that in actual theories one considers
\emph{scaled} fields (\sref{s:Extended electroweak geometry and fields}).

\begin{remark}
While we have isomorphisms
\hbox{$\FRa\cong\FR$} and  \hbox{$\FRa\cong\FR$}\,,
the 2-spinor bundle $\U$ is \emph{not} endowed with an analogous structure.
Hence \hbox{$\Ya\ncong\Y$}.\end{remark}

\subsection{Symmetry breaking}
\label{ss:Symmetry breaking}

As an added feature of the underlying geometric structure
we consider a fixed section
$$\Hivac:\M\to\FL\tn\FRl~,$$
called the ``vacuum value'' of the Higgs field.
This determines a splitting
$$\FL=\FR'\dir{\M}\FR^\sbot~,\quad \FR'\equiv\Hivac(\FR)~.$$
Let's assume that $\Hivac$ is of maximal rank $\dim\FR$\,,
namely that $\FR$ is isomorphic to its image
\hbox{$\FR'\equiv\Hivac(\FR)\subset\FL$}\,.
Then the fermion field can be decomposed as
\begin{align*}
&\Psi\equiv(\Psi_\Rrr\,,\,\Psi_\Lll)
=(\Psi_\Rrr\,,\,\Psi'_\Rrr\,,\,\Psi_\Rrr^\sbot)\equiv(\psi\,,\,\n)~,
\\[6pt]
&\psi\equiv(\Psi_\Rrr\,,\,\Psi'_\Rrr):
\M\to (\FR\tn\U)\oplus(\FR'\tn\Ua)\cong\FR\tn\W~,
\\[6pt]
&\n\equiv\Psi_\Rrr^\sbot:\M\to\FR'\tn\Ua\subset\FL\tn\Ua~.
\end{align*}

It's also natural to assume that $\Hivac$ has the further property of being
conformally isometric, namely
$$h_\Lll\comp(\bHivac,\Hivac)=
\frac{\m^2}{\scriptstyle\dim\FR}\,h_\Rrr~,\quad \m\in\RR~,$$
where \hbox{$h_\Lll:\M\to\FLa\tn\FLl$}
and \hbox{$h_\Rrr:\M\to\FRa\tn\FRl$}
denote the Hermitian metrics of $\FR$ and $\FL$\,.
We note that this condition implies
\hbox{$\bang{\bHivac,\Hivac}=\m^2$},
so that $\Hivac$ is a minimum of the ``Higgs potential''
\hbox{$\l\,(2\,\m^2\,\bang{\bar\phi,\phi}-\bang{\bar\phi,\phi}^2)$}\,,
\hbox{$\l\in\RR^+$}.

The $\Hivac$-splitting of $\FL$\,, together with the metric $h_\Lll$\,,
yields a splitting \hbox{$\FLl=\FR'{}^\lin\oplus\FR^{\sbot\lin}$},
so that
\begin{align*}
\Lie_\Lll\subset\End\FL&=
(\FR'\oplus\FR^\sbot)\tn(\FR'{}^\lin\oplus\FR^{\sbot\lin})=
\\[6pt]
&=(\FR'\tn\FR'{}^\lin)
\oplus(\FR^\sbot\tn\FR'^\lin)
\oplus(\FR'\tn\FR^{\sbot\lin})
\oplus(\FR^\sbot\tn\FR^{\sbot\lin})~.
\end{align*} 
Now consider the decomposition of any \hbox{$\xi\in\Lie_\Lll$} as
$$\xi=\xi'+\xi^{+}+\xi^{-}+\xi^\sbot
\in\Lie_\Rrr'\oplus\Lie_\Rrr^{+}\oplus\Lie_\Rrr^{-}\oplus\Lie_\Rrr^\sbot~,$$
where 
\begin{align*}
&\Lie_\Rrr'=\mathfrak{u}(\FR')\subset\FR'\tn\FR'^\lin~,
&&\Lie_\Rrr^\sbot=\mathfrak{u}(\FR^\sbot)\subset\FR^\sbot\tn\FR^{\sbot\lin}~,
\\[6pt]
&\Lie_\Rrr^{+}\equiv\FR^\sbot\tn\FR'^\lin~,
&&\Lie_\Rrr^{-}\equiv\FR'\tn\FR^{\sbot\lin}~.
\end{align*}
We easily realize that $\Lie_\Rrr^{+}$ and $\Lie_\Rrr^{-}$ are anti-isomorphic
by Hermitian adjunction,
and any \hbox{$\xi\in\Lie_\Lll$} (being anti-Hermitian)
fulfills \hbox{$(\xi^{-})^\dag=-\xi^{+}$}.
Hence we get a splitting
$$\Lie_\Lll\cong
\Lie_\Rrr'\oplus\Lie_\Rrr^{+}\oplus\Lie_\Rrr^\sbot\cong
\Lie_\Rrr'\oplus\Lie_\Rrr^{-}\oplus\Lie_\Rrr^\sbot~.$$

\smallbreak
Accordingly, symmetry breaking yields decompositions
of the left sector gauge field,
and of the ghost and anti-ghost sectors.

\section{Extended electroweak geometry and fields}
\label{s:Extended electroweak geometry and fields}
\subsection{Standard electroweak geometry rivisited}
\label{ss:Standard electroweak geometry rivisited}

Electroweak geometry can be seen~\cite{C10a} as a specialization of the scheme
presented in~\sref{ss:Two-spinors and non-abelian gauge theories},
where $\FR$ and $\FL$ are both constructed from one main ingredient:
a complex vector bundle $\I\to\M$, called the \emph{isospin} bundle,
whose $2$-dimensional fibers are endowed with a Hermitian metric $h$\,.
Namely we set \hbox{$\FR\equiv\weu2\I$} and \hbox{$\FL\equiv\I$}\,,
so that the fermion bundle is
$$\Y\equiv\Y\!_{\!\Rrr}\oplus\Y\!_{\!\Lll}\equiv
\bigl(\weu2\I\tn\U\bigr)\oplus\bigl(\I\tn\Ua\bigr)~.$$

With respect to said general scheme we'll now consider scaled fields:
the fermion field
$$\Psi\equiv\Psi_\Rrr\,{+}\,\Psi_\Lll:
\M\to\LL^{-3/2}\tn(\Y\!_{\!\Rrr}\,{\oplus}\,\Y\!_{\!\Lll})~,$$
gauge fields
\begin{align*}
&X\equiv A_\Rrr:\M\to\LL^{-1}\tn\H^*\tn\Lie_\Rrr~,&&
W\equiv A_\Lll:\M\to\LL^{-1}\tn\H^*\tn\Lie_\Lll~,
\\[6pt]
&\Lie_\Rrr\subset \weu2\I\tn\weu2\Il~,&&
\Lie_\Lll\subset \I\tn\Il~,
\end{align*}
and the Higgs and anti-Higgs fields
\begin{align*}
&\phi:\M\to\LL^{-1}\tn\weu2\Ic\tn\I\cong\LL^{-1}\tn\I\tn\weu2\Il~,
\\[6pt]
&\bar\phi:\M\to\LL^{-1}\tn\weu2\I\tn\Ic\cong\LL^{-1}\tn\weu2\I\tn\Il~.
\end{align*}

\begin{remark}
In the usual presentations of the electroweak theory
the left and right sectors are linked through a ``charge formula'' which,
in practice, determines the relation between the covariant derivatives
in these sectors.
This relation can be exactly recovered by the assumption
\hbox{$\FR\equiv\weu2\I$}, \hbox{$\FL\equiv\I$},
if we use the connection of $\weu2\I$ naturally determined
by the connection of $\I$.
As for the relation between the fields $X$ and $W$ above,
we note that the 
Hermitian metric of $\I$
determines the Hermitian metric of $\weu2\I$,
and any anti-Hermitian endomorphism $\Xi$ of $\I$
determines an anti-Hermitian endomorphism $\hat\Xi$ of $\weu2\I$.
Hence we may consider, in particular,
gauge fields such that \hbox{$A_\Rrr\equiv\hat A_\Lll$}\,.
Furthermore, we note that the fibers of $\Lie_\Rrr$
are isomorphic to $\iO\RR$\,,
because the fibers of $\weu2\I$ are 1-dimensional.

\end{remark}\smallbreak

Let $\bigl(\xi_\a\bigr)$\,, \hbox{$\a=1,2$}\,,
be an $h$-orthonormal local frame of $\I\onto\M$ (\emph{isospin frame}),
and $\bigl(\xi^\a\bigr)$ its dual frame.
We have the induced frames \hbox{$\hat\xi\equiv\xi_1\we\xi_2$}
of \hbox{$\FR\cong\weu2\I$},
\hbox{$\hat\xi{}^*\equiv\xi^1\we\xi^2$} of $\weu2\Il$, and
$$\bigl(\io_\m\bigr)\equiv\bigl(\s\iIi\m\a\b\,\xi_\a\tn\xi^\b\bigr)$$
of $\iO\Lie_\Lll$\,, where $\bigl(\s_\m\bigr)$ are the Pauli matrices.
Accordingly, we write the fields' coordinate expressions as\footnote{%
The constant $q$ is useful for a closer comparison
with the formulas found in the literature,
where it is usually denoted as $g$\,.} 
\begin{align*}
&\Psi=\Psi^\sA\,\hat\xi\tn\ze_\sA+\Psi^\a_\cA\,\xi_\a\tn\bze^\cA~,
\\[6pt]
&X=\iO\,q\,X_\l\,\t^\l~,\quad
W=\ih\,q\,W_\l^\m\,\io_\m\tn\t^\l~,\quad q\in\RR^+~,
\\[6pt]
&\phi=\phi^\a\,\xi_\a\tn\hat\xi{}^*~,\quad
\bar\phi=\phi_\a\,\hat\xi\tn\xi^\a~,
\end{align*}
where $\bigl(\ze_\sA\bigr)$ and $\bigl(\t_\l\bigr)$
are a two-spinor frame and the related Pauli frame
(\sref{ss:Two-spinors and Einstein-Cartan-Maxwell-Dirac field theory}).
The scaling is carried by the field's components.
Note that the condition \hbox{$X=\hat W$} reads \hbox{$X_\l=W_\l^0$}.
Furthermore, note that $W_\l^\m\,\io_\m$ is Hermitian,
as $W$ is anti-Hermitian valued by definition.

\subsection{EW symmetry breaking and standard Higgs interactions}
\label{ss:EW symmetry breaking and standard Higgs interactions}

The special section \hbox{$\Hivac:\M\to\LL^{-1}\tn\I\tn\weu2\Il$}
is a minimum of the `Higgs potential'
$$V[\phi]:=\l\,(2\,m^2\,\bang{\bar\phi,\phi}-\bang{\bar\phi,\phi}^2)~,$$
with \hbox{$m\in\LL^{-1}$} and \hbox{$\l\in\RR^+$}.
This determines an $h$-orthogonal decomposition $\I=\I_{\!1}\oplus\I_{\!2}$
characterized by \hbox{$\Hivac:\M\to\LL^{-1}\tn\I_{\!1}\tn\weu2\Il$}\,.
We can choose the $h$-orthonormal isospin frame $\bigl(\xi_\a\bigr)$
in such a way that $\Hivac=m\,\xi_1\tn\hat\xi{}^*$.
Then we write the coordinate expressions
\begin{align*}
\phi&=(m+\reH+\iO\,\phizero)\,\xi_1\tn\hat\xi{}^*
+\phiplus\,\xi_2\tn\hat\xi{}^*~,
\\[8pt]
\bar\phi&=(m+\reH-\iO\,\phizero)\,\hat\xi\tn\xi^1
+\phiminus\,\hat\xi\tn\xi^2~,\quad\phiminus\equiv\overline{\phiplus}~,
\end{align*}
where the \emph{Higgs ghosts}
$\phizero$ and $\phi_{\pm}$ are respectively real and complex.

We'll use the shorthand \hbox{$\E\equiv\End\I\cong\I\tn\Il$}.
Symmetry breaking,
together with the choice of a new constant $\th\in(0,\pi/2)$
(the \emph{Weinberg angle}),
determines the decomposition
\hbox{$\E=\E'\oplus\E''\oplus\E_{+}\oplus\E_{-}$}\,, where
$$\E'\equiv\I_{\!1}\tn\Il_{\!1}~,\quad
\E_{+}\equiv\I_{\!2}\tn\Il_{\!1}~,\quad
\E_{-}\equiv\I_{\!1}\tn\Il_{\!2}~,\quad~,$$
and \hbox{$\E''\subset(\I_{\!1}\tn\Il_{\!1})\oplus(\I_{\!2}\tn\Il_{\!2})$}
is generated by
$$\ee''\equiv -\sin\th\,\tan\th\,\io_0+\cos\th\,\io_3=
\sec\th\,\bigl[\cos(2\th)\,\xi_1\tn\xi^1-\xi_2\tn\xi^2\bigr]~.$$
Together with $\ee''$\,, the sections
\begin{align*}
&\ee'\equiv-\sin\th\,(\io_0+\io_3)=-2\,\sin\th\,\xi_1\tn\xi^1~,
\\[6pt]
&\ee^{+}\equiv\osq\,(\io_1-\iO\,\io_2)=\sqrt2\,\xi_2\tn\xi^1~,\quad
\ee^{-}\equiv\osq\,(\io_1+\iO\,\io_2)=\sqrt2\,\xi_1\tn\xi^2~,
\end{align*}
constitute a (not orthogonal) frame of $\E$ adapted to the above splitting.
We then write the gauge field in the left-handed sector
as \hbox{$W=\ih\,q\,\t^\l\tn W_\l$} with
$$W=\ih\,q\,\t^\l\tn\bigl(W_\l^\m\,\io_\m\bigr)=
\ih\,q\,\t^\l\tn\bigl(
A_\l\,\ee'+Z_\l\,\ee''+W^{+}_\l\,\ee^{+}+W^{-}_\l\,\ee^{-}\bigr)~.$$

A Fermion field splits as
\hbox{$\Psi\equiv\Psi_\Rrr+\Psi_\Rrr'+\Psi_\Rrr^\sbo\equiv\psi+\n$}\,;
\hbox{$\psi\equiv\Psi_\Rrr+\Psi_\Rrr'$} is the electron field
and \hbox{$\n\equiv\Psi_\Rrr^\sbo$} is the neutrino.

The standard procedure for determining the interactions
consists in viewing gauge fields as connections,
and examining a Lagrangian density written in terms
of curvature tensors and covariant derivatives of the matter fields.
The Higgs field interacts with the fermion field through the term
\hbox{$-( \bang{\bar\Psi_\Lll\,\phi\,\Psi_\Rrr}
+\bang{\bar\Psi_\Rrr\,\bar\phi\,\Psi_\Lll})\,\rdg\,\dO^4x$}\,;
its interactions with the gauge fields and itself 
are extracted from the `Higgs Lagrangian'
\hbox{$\Lcal_\phi=\ell_\phi\,\dO^4x$}\,, where
$$\ell_\phi=\bigl(g^{\l\m}\,\na_\l\bar\phi_\a\,\na_\m\phi^\a
+V[\phi]\bigr)\,|\det g|^{1/2}~.$$
Here $g$ is the spacetime metric,
and $\nabla\phi$ must be intended as covariant derivative
with respect to a connection \hbox{$\G_{\!0}+W$} where the `gauge' $\G_{\!0}$
is a locally chosen flat connection
(\sref{ss:Remarks about classical and quantum gauge theories}).
We get \hbox{$\nabla\phi=\na_\l\phi^\a\,\dO x^\l\tn\xi_\a\tn\hat\xi{}^*$} with
\begin{align*}
\na_\l\phi^1&=
\de_\l\reH+\iO\,\de_\l\phizero
-\ih\,q\,\sec\th\,(m+\reH+\iO\,\phizero)\,Z_\l
-\isq\,q\,\phiplus\,W^{-}_\l~,
\\[6pt]
\na_\l\phi^2&=\de_\l\phiplus
-\iO\,\sin\th\,q\,\phiplus\,A_\l
+\ih\,\sec\th\,\cos(2\th)\,q\,\phiplus\,Z_\l \\
&\hspace{5cm}
-\isq\,q\,(m+\reH+\iO\,\phizero)\,W^{+}_\l~.
\end{align*}
Let's now make the replacements \hbox{$\de_\l\reH\to\iO\,p_\l\,\reH$}
and the like, where $p$ is the appropriate 4-momentum.
Then from $\ell_\phi$ we indeed get the standard Higgs interactions
as listed for example in Veltman~\cite{Ve}, Appendix~E.2
(allow for different conventions).
Similarly, we can recover all interactions of the electroweak theory~\cite{C10a}.

\subsection{Further scalar invariants from Higgs geometry}
\label{ss:Further scalar invariants from Higgs geometry}

Accepting the idea about gauge fields discussed
in~\sref{ss:Remarks about classical and quantum gauge theories},
and treating the gauge fields of electroweak theory
as sections \hbox{$\M\to\H^*\tn\I\tn\Il$}
rather than connections of \hbox{$\I\onto\M$},
we find several more scalars and, consequently, point interactions,
than are derived from the usual Lagrangian.
The full gauge invariance of the Lagrangian is broken by such terms,
but is preserved by fermion interactions,
which, at least in the standard theory, determine 
the relation between gauge fields and connections.
Speculating further, a possible breaking of the bond between gauge field
and classical connection, in the extended theory,
would not be necessarily to rule out,
and could have interesting physical consequences,
though of course various issues could arise in this scenario,
in particular with regard to renormalization.\footnote{%
For a deeper examination of the physical consequences one may
also consider possible extensions of the ghost Lagrangian,
for example by adding terms such as one obtains
by various contractions of quantities of the type
$\gh\tn\agh\tn\phi\tn\bar\phi$\,.} %

Consider the tensor field $W\tn W\tn\phi\tn\bar\phi$\,,
which has the component expression\footnote{%
Since $\iO W$ is Hermitian-valued,
$W$ and $\bar W$ can be seen, in practice, as the same field.
In this expression primed indices are just regular isospin indices:
the primes allow us not to use too many greek characters.} %
$$W\iIi\l\a{\a'}\,W\iIi\m\b{\b'}\,\phi^\g\,\bar\phi_{\g'}~.$$
Hence the contraction
$$\Ical_1\equiv g^{\l\m}\,W\iIi\l\a{\a'}\,W\iIi\m\b\a\,\phi^{\a'}\,\bar\phi_\b$$
is now an invariant (and also a term in $\ell_\phi$).
Since we view $W$ as an unconstrained tensor field,
we can obtain more scalars by isospin index permutations in $\Ical_1$\,.
Explicit calculations show that there are only three distinct such scalars,
namely
$$g^{\l\m}\,W\iIi\l\a{\a'}\,W\iIi\m\b\b\,\phi^{\a'}\,\bar\phi_\a~,\quad
g^{\l\m}\,W\iIi\l\a\a\,W\iIi\m\b\b\,\phi^\g\,\bar\phi_\g~,\quad
g^{\l\m}\,W\iIi\l\a\b\,W\iIi\m\b\a\,\phi^\g\,\bar\phi_\g~,$$
which we denote respectively as $\Ical_2$\,, $\Ical_3$ and $\Ical_4$
(for brevity we do not write down their explicit expressions).
Moreover they are not all independent, as a straightforward calculations yields
$$2\,\Ical_1-2\,\Ical_2+\Ical_3-\Ical_4=0~.$$

We obtain still more invariants by considering the complex `symplectic' structure
in the fibers of $\I$,
analogous to the 2-form $\e$ considered
(\sref{ss:Two-spinors and Einstein-Cartan-Maxwell-Dirac field theory})
for the 2-spinor bundle,
and denoted here, for simplicity, by the same symbol.
We find three distinct invariants, namely
\begin{align*}
&\Jcal_1\equiv g^{\l\m}\e_{\a\b}\,\e^{\a'\b'}
W\iIi\l\a{\a'}\,W\iIi\m\b{\b'}\,\phi^\g\bar\phi_\g\,,~
\Jcal_2\equiv g^{\l\m}\e_{\a\g}\,\e^{\a'\g'}
W\iIi\l\a{\a'}\,W\iIi\l\b\b\,\phi^\g\bar\phi_{\g'}\,,
\\[6pt]
&\Jcal_3\equiv g^{\l\m}\e_{\a\g}\,\e^{\b'\g'}
W\iIi\l\a\b\,W\iIi\l\b{\b'}\,\phi^\g\bar\phi_{\g'}\,.
\end{align*}
These turn out to be not all independent, too, as we find
$$\Jcal_1-2\,\Jcal_2+2\,\Jcal_3=0~.$$

By a careful examination one can also show that there is no further
non-trivial vanishing linear combination
\hbox{$\sum_i x_i\,\Ical_i+\sum_j y_j\,\Jcal_j$}\,.

\subsection{Higgs potential revisited}
\label{ss:Higgs potential revisited}

Recalling~\sref{ss:Two-spinors and non-abelian gauge theories}
and~\sref{ss:Standard electroweak geometry rivisited}
we see how the sectors of standard electroweak geometry
can be recovered as sectors of $\Yc\tn\Y$.
Following our point of view
we now look for further sectors by considering $\Y\tn\Yl$.
Besides ghost and anti-ghost sector, we also find
$$\FL\tn\FRl\tn\Ul\tn\Ua\equiv\FL\tn\FRl\tn\H^*~.$$
In the electroweak case we then consider fields
$$\FH:\M\to\I\tn\weu2\Il\tn\H^*~,\quad
\bar\FH:\M\to\weu2\I\tn\Il\tn\H^*~.$$
Note how these are analogous to the Higgs field
in that they mix the right-handed and left-handed sectors,
but they are actually \hbox{spin-1} fields.
As usual we denote 2-spinor indices by Latin capitals,
and conjugated indices as dotted indices,
so that the above fields' components are written as $\FH^\a_\AAd$ and $\bar\FH_{\a\AAd}$\,,
or as $\FH^\a_\l$ and $\bar\FH_{\a\l}$ when we use spacetime indices $\l,\m\,$...
We remark that, even if the components have three indices,
these actually represent particles with four `chemical bonds',
the fourth being related to the 1-dimensional fiber type of $\weu2\I$.

We now aim at seeing in which ways, from these fields,
one can form scalars that can be possibly added to the total Lagrangian.
Our first observation is that if \hbox{$m\in\LL^{-1}$}
then $m\,(\FH,\bar\FH)$ can be seen,
via the tetrad, as a 1-form valued into the endomorphisms of
\hbox{$\FR\dir{\M}\FL\equiv\weu2\I\dir{\M}\I$}\,,
as well as $(A_\Rrr\,,A_\Lll)$\,.
Hence we may use $m\,(\FH,\bar\FH)$ in order to modify
the covariant derivative of the Higgs field in the Higgs Lagrangian
$g^{\l\m}\,\na_\l\bar\phi_\a\,\na_\m\phi^\a$,
obtaining non-kinetic terms such as
$-m^2\,g^{\l\m}\,\bar\phi_\a\,\phi^\b\,\bar\FH_{\b\l}\,\FH^\a_\m$
and
$-m^2\,g^{\l\m}\,\bar\phi_\a\,\phi^\a\,\bar\FH_{\b\l}\,\FH^\b_\m$~.
However we wish to explore the various possibilities a little more systematically,
beginning with considering isospin and spacetime index contractions
(later we'll enlarge our list by considering 2-spinor index contractions).
Furthermore we'll consider the possibility that $\FH$ be either of bosonic type
or of fermionic type;
actually, though the usual Lagrangians are essentially unchanged
when one assumes the fields to be valued into the $\ZZ_2$-graded operator algebra,
the situation we are exploring may turn out to be slightly more intricate in some cases.

We use the shorthands
\begin{align*}
&M\iI{\l\m\a}\b\equiv m^2\,\bar\FH_{\l\a}\,\FH\iI\m\b~,
&&M_{\l\m}\equiv M\iI{\l\m\a}\a~,
\\[6pt]
&M\iI\a\b\equiv g^{\l\m}\,M\iI{\l\m\a}\b~,
&&M\equiv M\iI\a\a=g^{\l\m}\,M_{\l\m}~,
\end{align*}
and note that
$$M\iI{\l\m\a}\b\,M\iI{\n\r\b}\a=
\pm M\iI{\l\r\a}\a\,M\iI{\n\m\b}\b\equiv
\pm M_{\l\r}\,M_{\n\m}$$
(the minus signs holds if $\FH$ and $\bar\FH$ are \emph{fermion} fields).
Considering all possible contractions we essentially obtain five distinct scalars
containing four $\FH$ factors:\footnote{%
Apparently there is a further possibility,
but actually a straightforward calculation yields\\
$g^{\l\r}\,g^{\m\n}\,\e^{\a\a'}\,\e_{\b\b'}\,
M\iI{\l\m\a}\b\,M\iI{\n\r\a'}{\b'}=\mp\Scal_4$\,.
} %
\begin{align*}
&\Scal_1\equiv g^{\l\m}\,g^{\n\r}\,M_{\l\m}\,M_{\n\r}\equiv M^2~,
\\[6pt]
&\Scal_2\equiv g^{\l\n}\,g^{\m\r}\,M_{\l\m}\,M_{\n\r}~,
\\[6pt]
&\Scal_3\equiv g^{\l\r}\,g^{\m\n}\,M_{\l\m}\,M_{\n\r}~,
\\[6pt]
&\Scal_4\equiv g^{\l\m}\,g^{\n\r}\,\e^{\a\a'}\e_{\b\b'}\,
M\iI{\l\m\a}\b\,M\iI{\n\r\a'}{\b'}~,
\\[6pt]
&\Scal_5\equiv g^{\l\n}\,g^{\m\r}\,\e^{\a\a'}\e_{\b\b'}\,
M\iI{\l\m\a}\b\,M\iI{\n\r\a'}{\b'}~.
\end{align*}
Here we used again (see~\sref{ss:Further scalar invariants from Higgs geometry})
the isospin ``symplectic'' form $\e$\,,
which is unique up to a phase factor so that
$\e^{\a\a'}\e_{\b\b'}$ is independent of it.

By a straightforward calculation we find the identity
$$\Scal_1\mp \Scal_3-\Scal_4=0~.$$
We also find that $\Scal_5$ vanishes in the bosonic case,
but not in the fermionic case,
in which one has the further identity
$$2\,\Scal_2-\Scal_5=0~.$$

It's interesting to observe that in the fermionic case
there are vanishing combinations of the five terms $\Scal_i$\,,
with integer coefficients all different from zero.
The simplest such combinations are
$$\Scal_1+\Scal_3-\Scal_4\pm(2\,\Scal_2-\Scal_5)~,$$
and one finds more combinations
by allowing greater coefficients.
In other terms, we could have a theory in which
the potentials of all point self-interactions of $\FH$ sum up to zero.
Extending our speculations we may suppose that this situation
triggers a symmetry breaking mechanism in which some non-zero vacuum value
of $\FH$ is selected.

What can then be said about the Higgs Lagrangian?
In the standard theory this is obtained
by adding the usual Higgs potential to the expression
$g^{\l\m}\,\na_\l\bar\phi_\a\na_\m\phi^\a$
containing the kinetic terms.
Now suppose we rather add the sum of all remaining 4-factor scalars
involving $\phi$ and $\FH$\,,
namely the term $-|\phi|^4$ minus the sum of the terms
\begin{align*}
&\Scal'_1\equiv m^2\,\FH^2\,|\phi|^2\equiv
m^2\,g^{\l\m}\,\bar\FH_{\l\a}\,\FH\iI\m\a\,\bar\phi_\b\,\phi^\b~,
\\[6pt]
&\Scal'_2\equiv g^{\l\m}\,M\iI{\l\m\a}\b\,\bar\phi_\b\,\phi^\a\equiv
m^2\,g^{\l\m}\,\bar\FH_{\l\a}\,\FH\iI\m\b\,\bar\phi_\b\,\phi^\a~,
\\[6pt]
&\Scal'_3\equiv g^{\l\m}\,\e^{\a\a'}\e_{\b\b'}\,
M\iI{\l\m\a}\b\,\bar\phi_{\a'}\,\phi^{\b'}\equiv
m^2\,g^{\l\m}\,\e^{\a\a'}\e_{\b\b'}\,
\bar\FH_{\l\a}\,\FH\iI\m\b\,\bar\phi_{\a'}\,\phi^{\b'}~.
\end{align*}
However, a straightforward calculation shows that actually
\hbox{$\Scal'_2+\Scal'_3=\Scal'_1$}
(this is true in both cases, bosonic and fermionic),
so that eventually we get a potential
$$-|\phi|^4-(\Scal'_1+\Scal'_2+\Scal'_3)=-(|\phi|^4+2\,m^2\,\FH^2\,|\phi|^2)~.$$
If the hypothesized vacuum value of $\FH$ is such that
\hbox{$\FH^2<0$} we then essentially recover the usual Higgs potential.

Next we explore the possibilities offered by 2-spinor index contractions.
These include all the above scalars, plus others.
Like above, isospin contractions can be performed essentially in two ways:
$$\bar\FH_{\a\AAd}\,\FH^\a_\BBd\,\bar\FH_{\b\CCd}\,\FH^\b_\DDd~,\quad
\e^{\a\a'}\e_{\b\b'}\,
\bar\FH_{\a\AAd}\,\FH^\a_\BBd\,\bar\FH_{\a'\CCd}\,\FH^{\b'}_\DDd~.$$
Multiplying each of these expressions 
by $\e^{\sA\sB}\,\e^{\sC\sD}\,\be^{\cA\cB}\,\be^{\cC\cD}$
we obtain 2-spinor index contractions,
actually all of them if we consider permutations
of the regular and conjugate indices separately;
essentially (taking the positive permutations) we consider products of
$$\e^{\sA\sB}\,\e^{\sC\sD}\,,~\e^{\sC\sA}\,\e^{\sB\sD}\,,
~\e^{\sA\sD}\,\e^{\sB\sC}~~\text{and}~~
\be^{\cA\cB}\,\be^{\cC\cD}\,,~\be^{\cC\cA}\,\be^{\cB\cD}\,,
~\be^{\cA\cD}\,\be^{\cB\cC}\,,$$
yielding a total of \hbox{$9+9=18$} scalars.

Now it can be checked, by straightforward calculations,
that the sum of each 9-uple of scalars identically vanishes;
moreover, this is true in the bosonic case and in the fermionic case as well.
Hence we find again, perhaps even more naturally, a situation in which
the potentials of all point self-interactions of $\FH$ sum up to zero.
The discussion about the recovery of the usual Higgs potential
then follows as above,
since 2-spinor contractions do not yield new terms from
$\bar\FH_\l\,\FH_\m\,\bar\phi\,\phi$\,.

\begin{remark}
The proposed mechanism yields a `breaking of dilatonic symmetry',
an issue which has been discussed by various authors~\cite{Fa,FKD,RySh,Ta}.
In a previous paper~\cite{C10a} I argued that these proposals
are essentially equivalent in the sense that they all
require an arbitrary choice of some value to be put in by hand.
While the mechanism proposed here does not determine such value,
it may help to explain how nature eventually chooses one.
\end{remark}\smallbreak

Finally, we might speculate that the above results be of some consequence
in a discussion of the problem of dark matter.

\subsection{Possible interactions of the extended Higgs sector}
\label{ss:Possible interactions of the extended Higgs sector}

In~\sref{ss:Two-spinors and non-abelian gauge theories}
we introduced the notion of an extended Higgs field
arising as a section of the sector \hbox{$\FL\tn\FRl\tn\End\Uc$},
which is one of the sectors of $\Yc\tn\Y$.
In the electroweak case, allowing for a necessary scaling,
we may consider sections
\begin{align*}
&\Phi:\M\to\LL^{-1}\tn\I\tn\weu2\Il\tn\Uc\tn\Ua~,
\\[6pt]
&\bar\Phi:\M\to\LL^{-1}\tn\weu2\I\tn\Il\tn\U\tn\Ul~.
\end{align*}
Within this setting we may identify the usual Higgs field as
\hbox{$\phi\equiv\Tr\Phi$}\,,
so that the extension is valued into the traceless endomorphisms of $\Uc$.

We have
$$\e^{\sB\sD}\,\e_{\sA\,\sC}\,\bar\Phi\iIi\a\sA\sB\,\bar\Phi\iIi\b\sC\sD=
\Tr\bar\Phi_\a\,\Tr\bar\Phi_\b-\Tr(\bar\Phi_\a\comp\bar\Phi_\b)$$
and the like,
hence we obtain just four distinct real contractions of four $\Phi$ factors, namely
\begin{align*}
&(\Tr\Phi^\a\Tr\bar\Phi_\a)^2~,
&& \Tr\Phi^\a\,\Tr\Phi^\b\,\Tr(\bar\Phi_\a\comp\bar\Phi_\b)~,
\\[6pt]
&\Tr\bar\Phi_\a\,\Tr\bar\Phi_\b\,\Tr(\Phi^\a\comp\Phi^\b)~,
&&\Tr(\Phi^\a\comp\Phi^\b)\,\Tr(\bar\Phi_\a\comp\bar\Phi_\b)~.
\end{align*}
Note that the first item in the above list
is essentially the quartic term of the usual Higgs potential.
Here, contractions via the isospin complex symplectic form vanish.

Next we consider scalars obtained as contractions of
$\bar\Phi\,\Phi\,\bar\FH\,\FH$\,.
Isospin index contractions yield the three distinct expressions
\begin{align*}
&\Phi\Ii{\a\cA}\cB\,\bar\Phi\iIi{\a}\sA\sB\,
\FH^\b_{\sC\cC}\,\bar\FH_{\b\sD\cD}~,\quad
\Phi\Ii{\a\cA}\cB\,\bar\Phi\iIi{\b}\sA\sB\,
\FH^\b_{\sC\cC}\,\bar\FH_{\a\sD\cD}~,
\\[6pt]
&\e^{\a'\b'}\e_{\a\b}\,
\Phi\Ii{\a\cA}\cB\,\bar\Phi\iIi{\a'}\sA\sB\,
\FH^\b_{\sC\cC}\,\bar\FH_{\b'\sD\cD}~,
\end{align*}
each of which yields nine scalars by 2-spinor index contractions.
In fact we may consider separate regular and dotted index contractions
via multiplications by
$$\d^\sB_\sA\,\e^{\sC\sD}\,,~\d^\sC_\sA\,\e^{\sD\sB}\,,~\d^\sD_\sA\,\e^{\sB\sC}
~~\text{and}~~
\d^\cB_\cA\,\be^{\cC\cD}\,,~
\d^\sC_\cA\,\be^{\cD\cB}\,,~\d^\cD_\cA\,\be^{\cB\cC}~.$$
Calculations show that, with certain choices of the signs,
the sum of the nine scalars may vanish in each of the three considered cases;
furthermore, one may have a situation in which the three sums do not vanish
but the overall sum does
(this is true if $\FH$ is either bosonic or fermionic).

Finally we point out that two-spinor index contractions
generate further possible three-leg interactions depending on momenta.
In the usual framework, point interactions can be directly recovered
in momentum representation
by replacing a partial derivative in the Lagrangian, say $\de_\l\phi^\a$\,,
by $\iO\,k_\l\,\phi^\a$ where $k$ is the momentum of $\phi$\,.
Indeed, this procedure yields all the standard interactions~\cite{C12a}.

The expression $\na_\l\bar\phi_\a\na_\m\phi^\a$
appearing in the standard Higgs Lagrangian
yields no further two-spinor index contractions than those contractions
obtained via multiplication by $g^{\l\m}$.
This is not true in the present context
in which we consider an extended Higgs sector,
as it can be seen by the following example.
Expanding $\na_\l\bar\Phi_\a\tn\na_\m\Phi^\a$ and using 2-spinor indices
one finds, in particular, terms of the type
$q\,W_\AAd\,k_\BBd\,\bar\Phi\iIi\a\cC\cD\,\Phi\Ii{\a\sC}\sD$\,.
Similarly to the above considered contractions of
$\bar\Phi\,\Phi\,\bar\FH\,\FH$\,,
we obtain nine scalars (not all independent) from this via multiplication by
$\d^\sA_\sC\,\e^{\sB\sD}\d^\cA_\cC\,\be^{\cB\cD}$
and separate permutations of the upper regular and dotted indices.



\begin{thebibliography}{AA}
%
\bibitem{LM}
M.\ Lavelle and D.\ McMullan:
`Observables and Gauge Fixing in Spontaneously Broken Gauge Theories'.\\
Phys.\ Lett.\ B {\bf 347} (1995), 89.
arXiv:9412145v1.
%
\bibitem{De02}
A.\ Derdzinski: 
`Geometry of the Standard Model',
notes of a talk at the Conference on Geometry
in Bedlewo, Poland, September 2002. 
%
\bibitem{Ta}
A.\ Talmadge:
`Symmetry breaking via internal geometry',
Int.\ J.\ Math.\ and Math.\ Sci.\ 2005:13 (2005), 2023--2030.
%
\bibitem{Fa}
L.D.\ Faddeev:
`An alternative interpretation of the Weinberg-Salam model',
Talk at the conference "New Trends in High Energy Physics", Yalta, Crimea, Sept.27-Oct.4 2008.
arXiv:hep-th/0811.3311v2.
%
\bibitem{FKD}
R.\ Foot, A.\ Kobakhidze and K.L.\ McDonald:
`Dilaton as the Higgs boson'.\\
arXiv:0812.1604v2.
%
\bibitem{RySh}
M.G.\ Ryskin and A.G.\ Shuvaev:
`Higgs Boson as a Dilaton'.\\
arXiv:0909.3374v1.
%
\bibitem{MT}
T.\ Masson and J.C.\ Wallet:
`A Remark on the Spontaneous Symmetry Breaking Mechanism
in the Standard Model'.
arXiv:1001.1176v1.
%
\bibitem{Mo10}
J.W.\ Moffat:
`Ultraviolet Complete Electroweak Model Without a Higgs Particle',
Eur.\ Phys.\ J.\ Plus 126:53 (2011);
arXiv:1006.1859v5.
%
\bibitem{NoBi}
M.\ Novello and E.\ Bittencourt:
`What is the origin of the mass of the Higgs boson?'
Phys. Rev. D {\bf 86}, 063510 (2012);
arXiv:1209.4871v1.
%
\bibitem{AMS}
O.\ Antipin, M.\ Mojaza and F.\ Sannino:
`Natural Conformal Extensions of the Standard Model',
Phys. Rev. D {\bf 89}, 085015 (2014);
arXiv:1310.0957v3.
%
\bibitem{C98}
D.\ Canarutto:
`Possibly degenerate tetrad gravity and Maxwell-Dirac fields',
J.\ Math.\ Phys.\ {\bf 39}, N.9 (1998), 4814--4823.
%
\bibitem{C00b}
D.\ Canarutto:
`Two-spinors, field theories and geometric optics in curved spacetime',
Acta Appl.\ Math.\ {\bf 62} N.2 (2000), 187--224.
%
\bibitem{C05}
D.\ Canarutto:
`Quantum bundles and quantum interactions',
Int.\ J.\ Geom.\ Met.\ Mod.\ Phys., {\bf 2} N.5, (2005), 895--917;
arXiv:math-ph/0506058v2.
%
\bibitem{C07}
D.\ Canarutto:
`{}``Minimal geometric data'' approach to
Dirac algebra, spinor groups and field theories',
Int.\ J.\ Geom.\ Met.\ Mod.\ Phys., {\bf 4} N.6, (2007), 1005--1040;
arXiv:math-ph/0703003.
%
\bibitem{C10a}
D.\ Canarutto:
`Tetrad gravity, electroweak geometry and conformal symmetry',
Int.\ J.\ Geom.\ Met.\ Mod.\ Phys., {\bf 8} N.4 (2011), 797--819;
arXiv:1009.2255v1 [math-ph].
%
\bibitem{C11a}
Canarutto: `Nature's software',
essay presented for the 2011 contest,
``Is Reality Digital or Analog?'',
of the Foundational Questions Institute (FQXi).\\
http://fqxi.org/community/forum/topic/831
%
\bibitem{C12a}
D.\ Canarutto: 
`Positive spaces, generalized semi-densities and quantum interactions',
J.\ Math.\ Phys. {\bf53} (3), 032302 (2012).\\
http://dx.doi.org/10.1063/1.3695348 (24 pages).
%
\bibitem{C14a}
D.\ Canarutto: 
`Two-spinor geometry and gauge freedom',
Int.\ J.\ Geom.\ Met.\ Mod.\ Phys., {\bf 11} (2014), 1460016 (18 pages);
arXiv:1404.5054v2 [math-ph].
%
\bibitem{C14b}
D.\ Canarutto: 
`Fr\"olicher-smooth geometries, quantum jet bundles and BRST symmetry',
J.\ Geom.\ Phys.\ (2014), http://dx.doi.org/10.1016/j.geomphys.2014.11.013.
arXiv:1405.1351 [math-ph].
%
\bibitem{JMV10}
J.\ Jany{\v s}ka, M.\ Modugno and R.\ Vitolo:
`An algebraic approach to physical scales',
Acta Appl.\ Math.\ {\bf 110} N.3 (2010), 1249--1276;
arXiv:0710.1313v1 [math.AC].
%
\bibitem{Penrose71}
R.\ Penrose:
`Angular momentum: an approach to combinatorial space-time',
in \emph{Quantum Theory and Beyond---%
essays and discussions arising from a colloquium},
Bastin T.\ editor, Cambridge Univ.\ Press, Cambridge (1971), 151--180.
%
\bibitem{Ve}
M.\ Veltman:
\emph{Diagrammatica},
Cambridge University press (1994).
%
\end{thebibliography}
\end{document}